\documentclass[conference]{IEEEtran}
\IEEEoverridecommandlockouts
\usepackage{cite}
\usepackage{amsmath,amssymb,amsfonts}
\usepackage{algorithmic}
\usepackage{graphicx}
\usepackage{textcomp}
\usepackage{xcolor}
\def\BibTeX{{\rm B\kern-.05em{\sc i\kern-.025em b}\kern-.08em
    T\kern-.1667em\lower.7ex\hbox{E}\kern-.125emX}}
\begin{document}

\title{Double Input Boost/Y-Source DC-DC Converter for Renewable Energy Sources\\
{\footnotesize \textsuperscript{*}}
\thanks{}
}

\author{\IEEEauthorblockN{Niteesh S Shanbog}
\IEEEauthorblockA{\textit{Dept. of Electrical and Electronics} \\
\textit{PES University}\\
Bengaluru, India \\
nit1995@yahoo.co.in}
\and
\IEEEauthorblockN{K R Pushpa}
\IEEEauthorblockA{\textit{Dept. of Electrical and Electronics} \\
\textit{PES University}\\
Bengaluru, India \\
pushpakr@pes.edu}
}

\maketitle

\begin{abstract}
With the increasing adoption of renewable energy sources by domestic users, decentralisation of the grid is fast becoming a reality. Distributed generation is an important part of a decentralised grid. This approach employs several small-scale technologies to produce electrical energy close to the end users or consumers. The higher reliability of these systems proves to be of advantage when compared to traditional generation systems. Multi-Input Converters (MICs) perform a decisive function in Distributed Energy Resources (DERs). Making use of such MICs prove to be beneficial in terms of size, cost, number of components used, efficiency and reliability as compared to using several independent converters. This thesis proposes a double input DC-DC converter which makes use of a quasi Y-source converter in tandem with a boost converter. The quasi Y-source converter has the advantage of having a very high gain for low duty cycles. The associated operating modes are analysed and the operation of the MIC is verified using simulation result. A hardware prototype is built for large signal analysis in open loop. Different loads are applied and the efficiency of the MIC as a whole as well as the load sharing between the different sources is investigated. 
\end{abstract}


\section{Introduction}
Multi Input Converters (MICs) are growing in popularity because of their new found applications in the modern energy system which includes significant amounts of renewable energy sources and distributed generation\cite{b9}. In most modern day setups, the input power source, output load demand or sometimes even both change instantaneously and are not exactly the same at any given instant\cite{b10} as they should ideally be. Providing a good match between the input source and the output sink is thus, a complicated task. Additional sources thus become necessary to aid the main source in load demand fulfillment.  

It is such situations that MIC’s have started to play a pivotal role. One single MIC can thus be able to replace numerous converters leading to significant savings. These converters interface different input voltage levels and combines their advantages to feed the load. MIC’s have a lower number of components while also simultaneously increasing the reliability and the efficiency. 

The Y-source impedance network is introduced in converters in order to achieve a high voltage gain even with small duty cycles. A DC-DC boost converter has been implemented in \cite{b1} using the Y-Source impedance network. This network makes use of a tightly coupled transformer to achieve the high gain.  
The quasi Y-Source DC- DC converter proposed in \cite{b2} is a modification to the Y-source converter. This modification over the Y-source converter enjoys continuous input current characteristics thus enabling it to be used in fuel cells and photo-voltaic systems. An entire family of Y-source based DC-DC converters have been defined in \cite{b11} to make use of it's high gain applications in sources like photo-voltaic cells and fuel cells. Coupled inductors have been used in literature\cite{b3} to create magnetically coupled impedance sources in converters to create high gain at small duty cycles. The Quasi Y-source converter belongs to this class of converters. The DC-DC Boost Converter\cite{b5} is one of the most basic power electronics converters which is used to boost the DC voltage. 

\begin{figure}[htbp]
\centering
\includegraphics[totalheight=0.17\textheight]{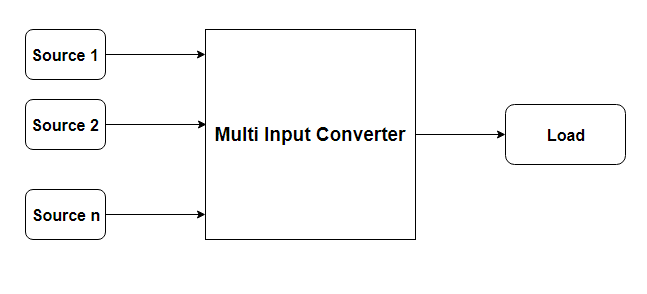}
\caption{General structure of a Multi Input Converter.}
\label{mic}
\end{figure}

Liu and Chen have given the general systematic approach to synthesize Multi Input Converters in \cite{b4}. The Quasi Y-source converter and the boost converter act like a Pulsating Current Source Cell according to \cite{b4} and hence they must be added together in parallel before being supplied to the load. Various types of double input converters which draw from two sources either simultaneously or individually have been implemented in \cite{b6},\cite{b7} and \cite{b8}.The general structure of an MIC is given in Fig~\ref{mic}.

This paper is organized into 7 sections. The next section contains the converter topology and the operating modes. Section III and IV discusses the simulation results and the observations from the hardware implementation respectively. Section V discusses about the results obtained and the pattern of load sharing between the two sources. Section VI discusses about the conclusions that can be drawn from this paper while Section VII talks about the future course of work that can be done with this converter topology. 

\section{Converter Topology}
\begin{figure}[htbp]
\centering
\includegraphics[totalheight=0.17\textheight]{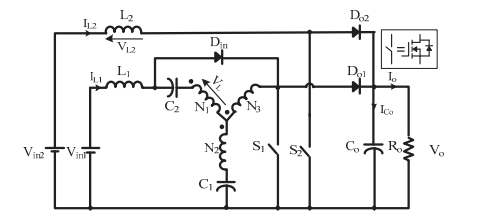}
\caption{General structure of a Multi Input Converter.}
\label{dic}
\end{figure}

The double input boost/Y-source DC-DC converter is using the guidelines given in \cite{b4}. Both the quasi Y-source converter and the boost converter act like a Pulsating Source Current (PCSC) cell when the load is removed. Hence they are parallely connected to the voltage sink in order to synthesize the Multi-Input Converter as shown in Fig~\ref{dic}. $V_{in1}$, $V_{in2}$ and $V_o$ are the input voltage of the quasi Y-source converter, the input voltage of the boost converter and output voltage at the load respectively. The source which is fed to the quasi Y-source converter has a smaller voltage than the second source because the quasi Y-source converter has a higher gain. 

$C_1$ and $C_2$ are  the capacitors added to the quasi Y-source converter in order to draw continuous input current. An additional input inductor, $L_1$ is added to stop the large inrush of input current at start-up and also to avoid input oscillations due to the input parasitic capacitances of $C_1$ and $C_2$.

The converter drains power from both the input sources either simultaneously or individually. Since there are two switches, we have four modes of operation:
\begin{itemize}
    \item \textit{Mode 1}: When SW1 is ON and SW2 is OFF. The load draws power only from Source 1. The input inductor of the second source is being charged. 
    \item \textit{Mode 2}: When both SW1 and SW2 is ON. The load is being supplied by both the power sources. 
    \item \textit{Mode 3}: When SW1 is OFF and SW2 is ON. The load is supplied only by the second source through the boost converter. The quasi Y-source converter is charging the input inductor. 
    \item \textit{Mode 4}: When both SW1 and SW2 is OFF. The load is completely disconnected from the source. 
\end{itemize}

The output equation of quasi Y-source converter is given in \cite{b2}: 
\begin{equation}
V_o = \dfrac{V_{in}}{1 - \delta d_{ST}} \label{eq}
\end{equation}

where $\delta$ is the an integral constant determined by the number of turns in the three winding transformer and is given by $\delta = \dfrac{N_1 + N_2}{N_2 - N_3}$ and $d_{ST}$ is the duty ratio for the quasi Y-source converter.

The output equation of a boost converter is given by 
\begin{equation}
    V_o = \dfrac{V_{in}}{1 - D} \label{eq}
\end{equation}
where $D$ is the duty cycle of the boost converter. Simluation is done on this converter using MATLAB-Simulink.  

\section{Simulation Results}
The double input converter has been simulated using MATLAB-Simulink. The circuit simulated is shown in Fig. 2 and the it's parameters are enumerated in Table 1.  
\begin{table}[htbp]
\caption{Circuit Parameters}
\begin{center}
\begin{tabular}{|c|c|}
\hline
\textbf{Parameters} & \textbf{Values} \\
\hline
Input Voltage ($V_{in1}$) & 12V  \\
\hline
Input Voltage ($V_{in2}$) & 24V \\
\hline
Output Voltage ($V_o$) & 48V \\
\hline
Input Inductors ($L_{in1}$ and $L_{in2}$) & 1mH \\
\hline
DC Blocking Capacitor ($C_1$) & $470\mu F$ \\
\hline
DC Blocking Capacitor ($C_2$) & $150\mu F$ \\
\hline
Output Capacitor ($C_o$) & $470\mu F$ \\
\hline
Turns Ratio of Three winding transformer ($N_1:N_2:N_3$) & 3:2:1 \\
\hline
Switching Frequency & $20kHz$\\
\hline
\end{tabular}
\label{tab1}
\end{center}
\end{table}

\begin{figure}[htbp]
\centering
\includegraphics[totalheight=0.19\textheight]{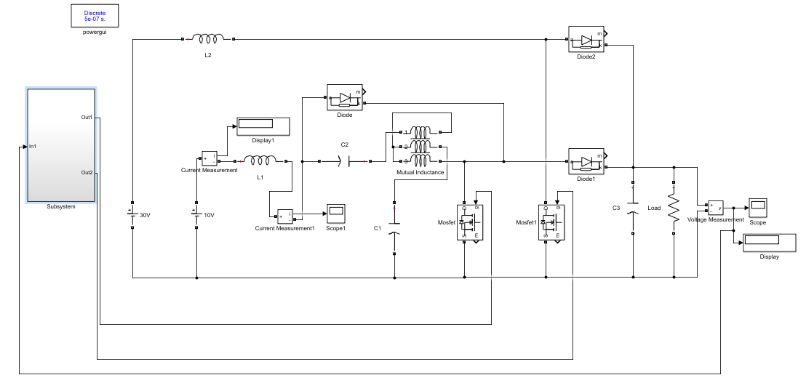}
\caption{Simulation Block Diagram of the Double Input Converter}
\label{sim}
\end{figure}

Fig~\ref{sim} shows the simulation block of the double input converter as used in MATLAB-Simulink. The turns ratio is selected from the required gain (4) in order to boost the input voltage from 12V to 48V. To get a gain of 4, the turns ratio is required to be 3:2:1.

Fig~\ref{out} shows the output voltage when the input voltages are 12V and 24V. 

\begin{figure}[htbp]
\centering
\includegraphics[scale=0.35]{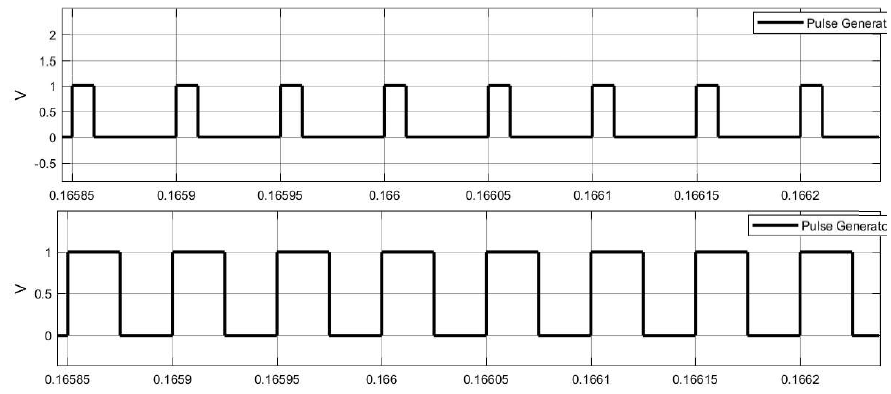}
\caption{Gate signals to the MOSFETs of Boost and Quasi Y-Source Converter}
\label{gate}
\end{figure}

\begin{figure}[htbp]
\centering
\includegraphics[totalheight=0.19\textheight]{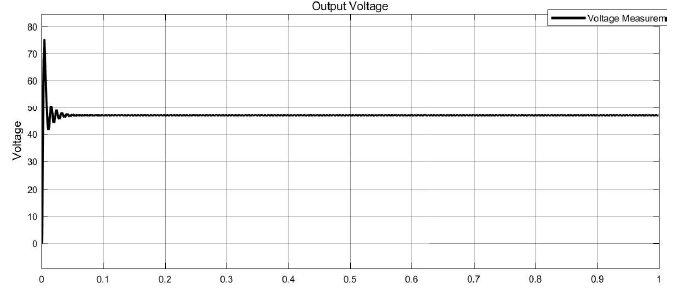}
\caption{Output Voltage of Double Input Converter}
\label{out}
\end{figure}

The gate signals to the MOSFETs of the two converters is shown in Fig~\ref{gate}. The Boost Converter requires a duty cycle of 50\% to boost an input voltage of 24V to 48V whereas the Quasi Y-Source Converter can boost 12V to 48V with a significantly lower duty cycle. 

Fig~\ref{out} shows the output voltage when the input voltages are 12V and 24V. 

\section {Hardware Implementation}
\begin{figure}[htbp]
\centering
\includegraphics[totalheight=0.22\textheight]{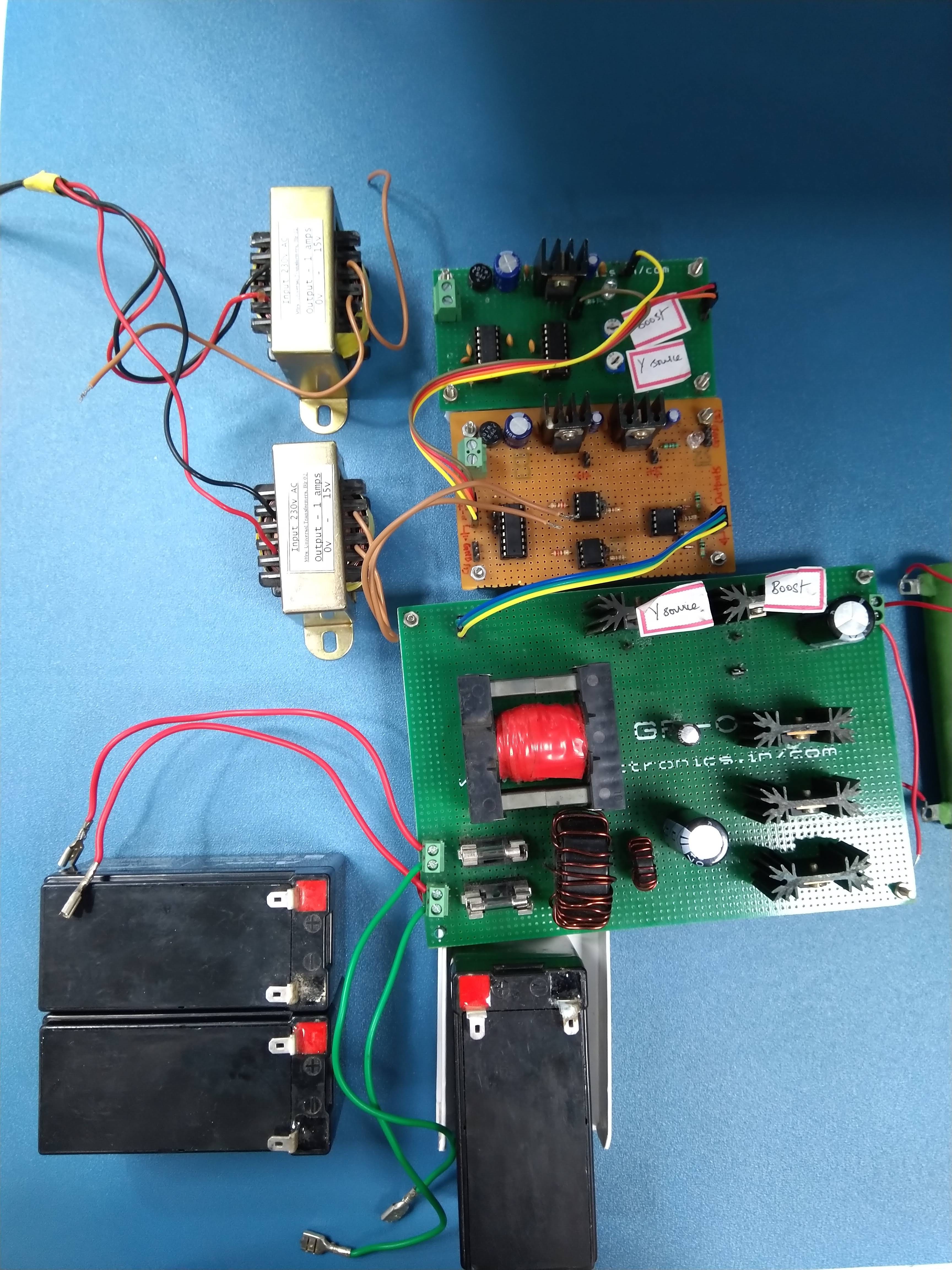}
\caption{Hardware model of the Double Input Converter}
\label{hard}
\end{figure}

An open loop implementation of the double input converter has been built and tested. The PWM signals are given by using two TL494 ICs in master slave configuration. This is to ensure that the PWM signals are synchronized. By inhibiting the internal oscillator of the slave TL494 and supplying the same saw-tooth waveform generated from the master to the timing capacitor ($C_t$) terminal of the slave TL494, synchronized PWM signals are possible. 
Fig~\ref{hard} shows the hardware of the double input converter which is built for testing purpose. 

IR2101 is used to drive the gates of the MOSFETs. The converter is loaded with different loads from 2.5W to 25W. The efficiency and the load sharing pattern of the converter is observed.

Two series of batteries are used as two independent sources. Two different voltage levels are used for this purpose. An output of 48V is obtained for inputs of 12V and 24V. 

\begin{figure}[htbp]
\centering
\includegraphics[totalheight=0.22\textheight]{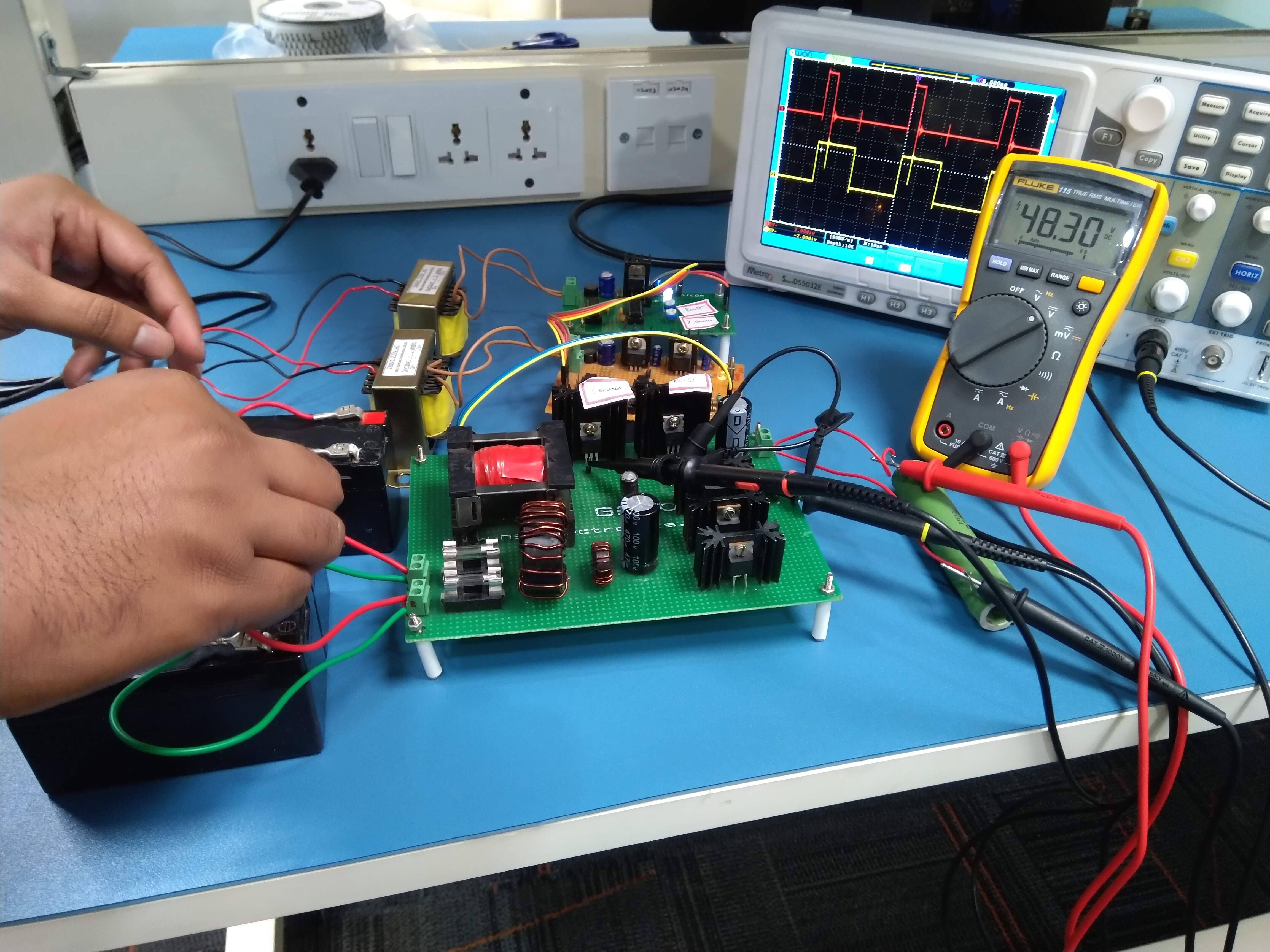}
\caption{Operation of the Double Input Converter}
\label{combihard}
\end{figure}

The same output is obtained even when each of the source is independently used as well as when they are used simultaneously.
Fig~\ref{combihard} shows the operation of the Double Input Converter when both the sources are connected to the two converters. 

Boost converters are inherently unstable as small increases in the duty cycle can lead to very high voltage gains. The tuning of the duty cycles is thus done beforehand. The output voltage of 48V is observed. 

\section{Results}
\begin{figure}[htbp]
\centering
\includegraphics[totalheight=0.23\textheight]{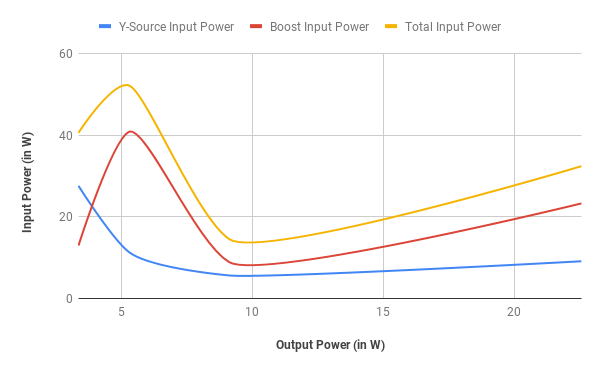}
\caption{Input Power vs Output Power}
\label{in-v-out-1}
\end{figure}

The trend of input power required to supply the given output load is shown in Fig~\ref{in-v-out-1}. It also shows the input power drawn by the two converters in order to supply the given load. 

\begin{figure}[htbp]
\centering
\includegraphics[totalheight=0.23\textheight]{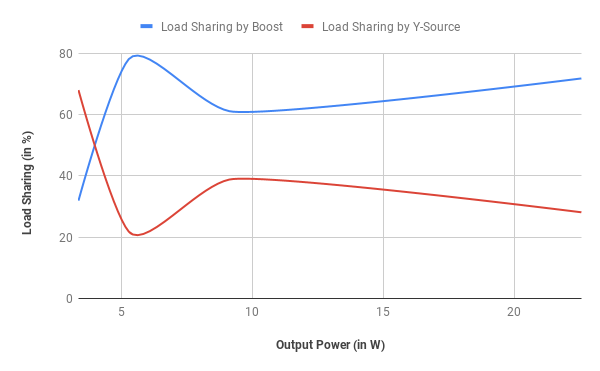}
\caption{Load Sharing between the two sources}
\label{load-sharing}
\end{figure}

This graph shows the pattern of load sharing by the two converters when different loads are applied. The efficiency of the converter is extremely low at small loads. The boost converter supplies the majority of the load in this converter in the open loop configuration. In the absence of any control strategy which dictates which source is to be the primary source, most of the current is hence drawn by the boost converter. This is because two 12V batteries are connected in series to derive the 24V input for the boost converter. Hence the strain on the battery is lower and this source can supply greater current than a single 12V battery. Fig~\ref{load-sharing} shows the sharing of load between the two converters at different values of loads.

Fig~\ref{load-sharing} shows that the boost converter which has 2 batteries supplies the majority of the load as the output load is increased. This is a potential problem because if one of the converter fails, then the other converter will go from supplying a percentage of the load to supplying the full load. During this time the voltage will droop in the absence of a closed loop control system. The transient will be even more severe if the converter which is supplying a higher percentage of the load fails. 

In the case of converters with multiple sources, most sources will not share the load equally. One of the sources will always try to supply an excessive fraction of the load. Typically, this is the source with the highest power rating. It will always try to deliver the highest current possible according to the design. The voltage will drop as it supplies more current and when the voltage has dropped to the point where it is lower than the next source, the other source will be activated. In a system with multiple sources, all the sources may contribute towards fulfilling the load, but the proportion is widely unequal. In the absence of any control strategy, the load sharing is highly dependent on the voltages of the different sources used.

\begin{figure}[htb]
		\begin{center}
		\includegraphics[scale=0.45]{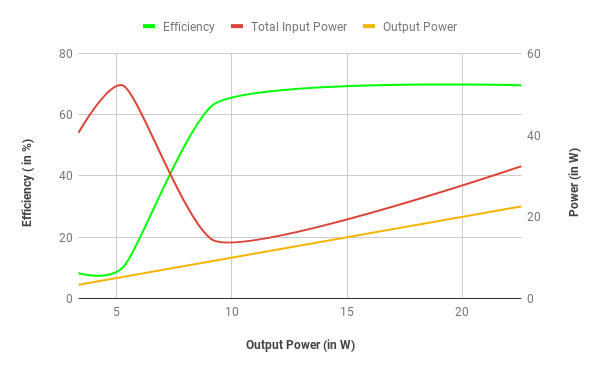}
		\caption{Efficiency of the Double Input Converter at different loads}
		\label{efficiency}
	\end{center}
	\end{figure}

Fig~\ref{efficiency} shows the plot of efficiency vs output power. By loading the double input converter with loads which are closer to the rated load, higher efficiency is obtained. At very low loads, the efficiency is very poor.  This can be explained by the conduction losses and the V-I overlap losses which are usually seen in DC-DC Converters at low loads. V-I overlap losses are proportional to input voltage, load current and switching frequency and occur due to the V-I overlap region that is seen in converters with rapid switching cycles.

But at low loads, the conduction losses which are caused due to current ripples dominate the overlap losses as the power dissipated due to current ripples remain same whereas the overlap losses reduce due to low load currents. Under light loads, the gate drive losses which occur due to the charging and discharging the gate capacitance dominate the load. 

\section{Conclusion}
With the increasing adoption of renewable energy in the power grid, it has become imperative to look for unique and efficient converter designs. Renewable energy sources are primarily DC in nature. It is also one of the leading types of distributed generation. In these conditions, the development of a Multi-Input Converter to aggregate all the different distributed sources and feed a common load assumes importance. This project proposes such a converter which can take in two different sources of different voltage levels, aggregate their power and feed a load. The use of different topologies of converters in this aggregated converter has been well demonstrated. A Quasi Y-Source converter which can give a high gain is used for this converter. A boost converter has also been used to demonstrate the use of multiple converters. With this converter, the load can be shared between the two sources. The use of multiple converters has been eliminated. This method of using a multiple input converter to aggregate different power sources can be used as a viable method to integrate renewables into the existing power infrastructure. 

\section{Future Scope}
To create a Multi Input Converter for distributed generation sources for real life applications, the open loop version will not be satisfactory. A control strategy has to developed where one source is chosen to be the primary source and supplies the bulk of the load while the other source acts as an auxiliary source kicking in at peak demands. The converter must be able to handle dynamic loads as well as dynamic sources as is the case with most of the renewable sources. Current sharing strategies must be developed so that both the converters supply a pre-determined share of the load. This is essential so that both the converters age at the same rate and replacement will be easier. 

An interesting method of current sharing will be to choose one converter as the master. This master converter will be employed with intelligent driver controls and will supply the base load. The slave controller will boost the voltage to meet the dynamic power requirements. 

The performance of the converter with different types of distributed generation sources like solar, wind, biomass, fuel cells etc must be evaluated. Primary sources must be selected based on the power availability. Auxiliary sources must be used to supply the peak load demand. Adding extra redundant converters to the existing topology can ensure reliability. Conducting reliability studies of such converters and developing redundancy indices will be an area of research.

Another upcoming are of research will be to develop bidirectional converters which allow the integration of storage devices as well. Advances in storage technology as well as the availability of cheap second hand batteries disposed from EV's will increase the use of batteries as a part of the power grid. Higher power production from renewables will be used to store the batteries which will then be used during peak loads. Having bidirectinal multi port converters will be useful in such conditions. 

\bibliographystyle{ieeetr}
\bibliography{bibliography.bib}

\vspace{12pt}

\end{document}